\documentclass[twocolumn,aps,pra,showpacs,superscriptaddress,nofootinbib,10pt]{revtex4-1}
\usepackage[latin1]{inputenc}
\usepackage{graphicx,dcolumn,bm}
\usepackage{subfigure}
\usepackage{color}
\usepackage[colorlinks,hyperindex]{hyperref}
\usepackage{amsfonts}
\usepackage{amssymb}
\usepackage{amsmath}
\usepackage{latexsym}
\usepackage{times,txfonts}
\usepackage{epsfig}

\usepackage{mathrsfs}
\usepackage{arydshln}
\usepackage[sans]{dsfont}

\definecolor{Blue}{rgb}{0.0,0.0,1}
\definecolor{Red}{rgb}{1,0.0,0.0}
\definecolor{Green}{rgb}{0,0.5,0.0}

\begin{document}

\title{Geometric lower bound for a quantum coherence measure}

\author{Diego Paiva Pires}
\email{diegopaivapires@gmail.com}
\affiliation{Instituto de F\'{i}sica de S\~{a}o Carlos, Universidade de S\~{a}o Paulo, P.O. Box 369, S\~{a}o Carlos, 13560-970 SP, Brazil}
\author{Lucas C. C\'{e}leri}
\affiliation{Instituto de F\'{i}sica, Universidade Federal de Goi\'{a}s, 74.001-970 Goi\^{a}nia, Goi\'{a}s, Brazil}
\author{Diogo O. Soares-Pinto}
\affiliation{Instituto de F\'{i}sica de S\~{a}o Carlos, Universidade de S\~{a}o Paulo, P.O. Box 369, S\~{a}o Carlos, 13560-970 SP, Brazil}
\begin{abstract}
Nowadays, geometric tools are being used to treat a huge class of problems of quantum information science. By understanding the interplay between the geometry of the state space and information-theoretic quantities, it is possible to obtain less trivial and more robust physical constraints on quantum systems. Here we establish a geometric lower bound for the Wigner-Yanase skew information (WYSI), a well-known information theoretic quantity recently recognized as a proper quantum coherence measure. In the case of a mixed state evolving under unitary dynamics generated by a given observable, the WYSI between the state and the observable is bounded from below by the rate of change of the state's statistical distinguishability from its initial value. Our result shows that, since WYSI fits in the class of Petz's metrics, this lower bound is the change rate of its respective geodesic distance on quantum state space. The geometric approach is advantageous because it raises several physical interpretations of this inequality under the same theoretical umbrella.
\end{abstract}


\maketitle


\section{Introduction}
\label{sec:sectionI}

Coherence is a striking feature of the quantum realm due to interference phenomena~\cite{Coherencebook_Ficek_Swain,*Coherencebook_Schlosshauer}. In fact, it is in equal footing with entanglement and other correlations whose meaning evades the classical view. Although quantum optics has proved to be a fruitful branch for quantum coherence studies~\cite{1996_PhysRevA.53.1295,*1999_PhysRevLett.83.5166,*2007_PhysRevLett.98.240401}, recent results suggest its connection with thermodynamics~\cite{2013_NatureCommun_4_2059,*2013_arxiv_1302.2811,*2013_arxiv_1308.1245,*2014_PhysRevLett.113.150402,*2014_arxiv_1405.5029,
*2014_arxiv_1410.4572v1} and quantum biology~\cite{2013_Nature_9_10,*Huelga_Plenio_10.1080_00405000.2013.829687,*2014_Nature_10_621}. Even some condensed matter phases, such as superconductivity and its emergent properties, display signatures of quantum coherence~\cite{2013_JPhysBAtomicMolOptPhys_10_104002}. As we know, the inability to perform basic tasks in quantum information processing is often related to coherence loss and, because of that, the interplay between noise and decoherence still holds as a key challenge in open quantum systems. Recently, Bromley and coworkers~\cite{2014_arXiv_14127161v1_BromleyCianciarusoAdesso} reported a possible way to circumvent this problem. Summarizing, they found a regime called {\it freezing conditions} in which coherence remains unchanged during the nonunitary dynamics.

Despite its fundamental role in physics, there is no unified way to characterize and quantify coherence. In consonance with results presented in Ref.~\cite{2008_NewJPhys_10_033023,*PhysRevA.80.012307}, a recent approach due to Baumgratz {\it et al}~\cite{PhysRevLett.113.140401} established a new paradigm in this scenario. By employing a rigorous mathematical framework for identifying proper coherence measures, they were able to classify {\it natural candidates} for coherence quantifiers based on distance measures, particularly relative entropy, $l_p$-norms, and fidelity. Simultaneously, Girolami~\cite{2014_arxiv_GirolamiDavide_14032446} proposed another quantum coherence measure based on the {\it Wigner-Yanase skew information} (WYSI), which shares the same reliable criteria of Ref. \cite{PhysRevLett.113.140401}. Besides the theoretical background, this work offers an efficient route to experimentally access the quantum coherence of an unknown state.

In the present work, we focus on skew information to provide an information-geometric lower bound for coherence measures. Introduced by Wigner and Yanase half century ago~\cite{1963_ProcNatlAcadSciUSA_49_910,*1964_CanadJMath_16_397}, skew information
\begin{equation}
 \label{eq:introduction001}
{\mathcal{I}}({\rho},\mathcal{K}) := -\frac{1}{2}\text{Tr}({[\sqrt{\rho},\mathcal{K}]^2})
\end{equation}
is a measure of the non-commutativity between a state ${\rho}$ and an observable $\mathcal{K}$. Operationally, this quantity is deeply more interesting than other coherence quantifiers because its calculation does not involve any optimization techniques. Also, it describes a constant of motion in closed quantum dynamics when the observable $\mathcal{K}$ is a conserved quantity, i.e., it commutes with the hamiltonian generating the evolution of the system~\cite{2003_PhysRevLett_180403}. Furthermore, WYSI is nonnegative, convex and vanishes if and only if the state and the observable commute~\cite{2004_IEEE_50_1778}. It is also bounded by the variance of $\mathcal{K}$, $\mathcal{I}(\rho,\mathcal{K}) \leq {\langle{\mathcal{K}^2}\rangle_{\rho}} - {\langle\mathcal{K}\rangle_{\rho}^2}$, an interesting property discovered by Luo which also noticed that the inequality is saturated for pure states~\cite{PhysRevA.73.022324}. This measure was later generalized by Dyson as
\begin{equation}
\label{eq:introduction002}
{\mathcal{I}^p}({\rho},\mathcal{K}) := -\frac{1}{2}\text{Tr}([{\rho^p},\mathcal{K}][{\rho^{1-p}},\mathcal{K}])
\end{equation}
with $0 < p < 1$, being called Wigner-Yanase-Dyson skew information (WYDSI), and its convexity proved by Lieb~\cite{1973_AdvMath_11_267,*1973_PhysRevLett_30_343}. 

There are several interpretations of the skew information, each one related to a particular viewpoint of the quantum behavior. Actually, the original one discusses the uncertainty in the measure of observables not commuting with a conserved quantity -- basically, the content of Wigner-Yanase-Araki theorem~\cite{PhysRev.120.622}. Similarly, WYSI supports a new type of Heisenberg uncertainty relation~\cite{2005_PhysRevA.72.042110,*2007_TheorMathPhys_151_693}, quantifies the quantum uncertainty of local observables~\cite{PhysRevLett.110.240402} and has applications in quantum reference frames and metrology~\cite{RevModPhys.79.555}. It is also possible to detect entangled states through a Bell-type inequality derived from the skew information~\cite{2005_PhysRevA_052302}.

WYSI is also an {\it asymmetry measure}, i.e., it quantifies symmetry breaking in a given state~\cite{PhysRevA.90.014102}. This is a promising subject in quantum information which finds support on the {\it asymmetry theory} and classifies coherence as a resource~\cite{2013_NewJPhys_033001}. In this context, Noether's theorem is a powerful tool to characterize conservation laws from symmetries in closed quantum systems because each asymmetry measure is a conserved quantity. Nevertheless, recent efforts have elucidated some asymmetry properties of pure states and quantum channels~\cite{2008_NewJPhys_10_033023,*PhysRevA.80.012307}, but the mixed state case is rather complex and less exploited. This happens because, when dealing with mixed states one must search for conservation laws which are not captured in its essence by Noether's theorem~\cite{Marvian_Spekkens_PhysRevA.90.062110}. As advocated by Marvian and Spekkens~\cite{2014_NatCommun_5_3821}, an asymmetry measure based on WYSI could fill this gap providing a way to point out more subtle features of conserved quantities.

The main result of our work is that, for closed quantum systems, the skew information, $\mathcal{I}({\rho_{\varphi}},{K_{\varphi}})$, between an evolved mixed state, $\rho_{\varphi}$, and the observable, $K_{\varphi}$, generating its evolution is lower bounded by the rate of change of the distinguishability between the evolved and the initial, $\rho_0$, mixed states
\begin{equation}
 \label{eq:wyhel0000}
\left|\frac{d}{d\varphi}\cos[\mathcal{L}({\rho_0},{\rho_{\varphi}})]\right| \leq
\frac{\sqrt{2}}{\hbar}\sqrt{\mathcal{I}({\rho_{\varphi}},{K_{\varphi}})}  ~.
\end{equation}
Here $\mathcal{L}({\rho_0},{\rho_{\varphi}})$ is the Hellinger angle between the initial state and the evolved state. The evolution is given by a family of unitary transformations $U_{\varphi}$ which changes continuously with respect to the parameter $\varphi$. The observable $K_{\varphi}$ may or not depend on 
the parameter $\varphi$ and it is connected with the operator $U_{\varphi}$ through the relation ${K_{\varphi}} = -i{\hbar}{U_{\varphi}}({dU_{\varphi}^{\dagger}}/{d\varphi})$. In our approach, the encoded parameter $\varphi$ can assume different interpretations depending on a specific physical situation. For instance, it could be the phase difference introduced in an interferometric protocol, with $K_{\varphi}$ being the generator of the rotation, or the time in a dynamical evolution, in which case $K_{\varphi}$ would be the Hamiltonian of the system.

The paper is organized as follows. In Sec.~\ref{sec:sectionII} we review the necessary and sufficient conditions that WYSI should satisfy in order to be a proper coherence measure. In Sec.~\ref{sec:sectionIII} we point out that this information-theoretic quantity defines a monotone Riemannian metric due to the Petz's theorem~\cite{1996_LinAlgApl_244_81} and its respective geodesic distance on quantum space state is given by the Hellinger angle~\cite{2003_JMathPhys_44_3752}. By exploring these two interfaces, in Sec.~\ref{sec:sectionIV} we demonstrate an inequality which assigns a geometric meaning to coherence measures. In other words, we show that, for closed quantum systems, the skew information between an evolved state and the observable generating the evolution is lower-bounded by the rate of change of the distinguishability between the evolved and the initial states of the system. In Sec.~\ref{sec:sectionV} we provide an example in order to illustrate our claim. Finally, in Sec.~\ref{sec:sectionVI}, we present our conclusions.


\section{WYSI and quantum coherence measures}
\label{sec:sectionII}

In order to characterize skew information as a coherence measure, it is essential to establish the concept of incoherent states and incoherent operations. An incoherent state is one that has no coherence, i.e., its off-diagonal elements are equal to zero. In the same way, an incoherent operation is one that does not create any kind of coherence. Despite the intuitive notions, in the following we will present these ideas in a rigorous fashion based on Refs.~\cite{PhysRevLett.113.140401,2014_arxiv_GirolamiDavide_14032446}.

It is well known that quantum operations are described by dynamical maps, i.e., quantum channels. In particular, the action of completely positive and trace preserving (CPTP) map $\mathcal{E}$ on the state $\rho$ can be synthesized as $\mathcal{E}(\rho) = {\sum_{\mu}}{K_{\mu}}\rho{K_{\mu}^{\dagger}}$, where $\{{K_{\mu}}\}$ is a set of Kraus operators satisfying ${\sum_{\mu}}{K_{\mu}^{\dagger}}{K_{\mu}} = I$. Let $\mathcal{H}$ be a finite-dimensional Hilbert space with $d = \dim\mathcal{H}$. Choosing a fixed basis $\{|i\rangle\}_{i = 1,\ldots,d}$, the subset of incoherent states $\mathscr{I} \subset \mathcal{H}$ encompasses those whose density matrix is diagonal in this basis. So, an incoherent channel (ICPTP) is the map ${K_{\mu}}\mathscr{I}{K_{\mu}^{\dagger}} \subset \mathscr{I}$ for all $\mu$, i.e., transform incoherent states into incoherent states. In other words, this constraint excludes any coherence generation process.

As demonstrated by Girolami~\cite{2014_arxiv_GirolamiDavide_14032446}, skew information is a faithful coherence measure since it satisfies the axiomatic postulates proposed by Baumgratz {\it et al}~\cite{PhysRevLett.113.140401}. First, it is convex, non-negative and vanishes for all incoherent states $\rho \in \mathscr{I}$. Indeed, $\mathcal{I}(\rho,\mathcal{K}) = 0$ if and only if $[\rho,\mathcal{K}] = 0$, i.e., state and observable can be diagonalized simultaneously. Secondly, it is monotonically nonincreasing under ICPTP maps and does not increase on average under a von Neumann measurement, $\mathcal{I}(\rho,\mathcal{K}) \geq {\sum_{\mu}}{p_{\mu}}\mathcal{I}({K_{\mu}}\rho{K_{\mu}^{\dagger}},\mathcal{K})$, where ${p_{\mu}} = \text{Tr}({K_{\mu}}\rho{K_{\mu}^{\dagger}})$.


\section{WYSI, Petz metric and Hellinger angle}
\label{sec:sectionIII}

WYSI is a robust information-theoretic quantifier due its enormous versatility. Actually, skew information also can be interpreted from a geometric perspective. The most remarkable approach to achieve this goal is indubitably due to Morozova-\v{C}encov~\cite{1990_Itogi_Nauki_i_Tehniki_36_69} and Petz~\cite{1996_LinAlgApl_244_81,1996_LettMathPhys_38_221,*2002_JPhysA_35_929,*2013_PhysRevA_87_032324}, by using monotone metrics on the quantum state space. In this space the set of density operators ($\rho \geq 0$ and $\text{Tr}\rho = 1$) constitute a differentiable manifold equipped with a suitable monotone Riemannian metric. By monotone metrics we consider the ones that are defined by positive, continuous and sesquilinear inner products which are also contractive under CPTP maps.

The Morozova-\v{C}encov-Petz theorem provides a friendly way to demonstrate that skew information fits in the category of monotone metrics and describes a particular kind of quantum Fisher information~\cite{1996_JMathPhys_37_2662}. Generally speaking, the theorem states that there exists a bijective correspondence between monotone metrics and operator monotone functions given by
\begin{equation}
 \label{eq:cencovmorozovfunc001}
{g_f}(A,B) := \text{Tr}[A \, {c_f}({\mathscr{L}},{\mathscr{R}}) \, B] ~,
\end{equation} 
where $A$ and $B$ are traceless hermitian operators and
\begin{equation}
 \label{eq:cencovmorozovfunc002}
{c_f}(x,y) := \frac{1}{yf(x/y)}
\end{equation}
is a symmetric function, ${c_f}(x,y) = {c_f}(y,x)$, and fulfils ${c_f}(\alpha x,\alpha y) = {\alpha^{-1}}{c_f}(x,y)$ with $x,y>0$. Here $\mathscr{L}\mathscr{O} = \rho \mathscr{O}$ and $\mathscr{R}\mathscr{O} = \mathscr{O}\rho$ are commuting operators, $[\mathscr{L},\mathscr{R}]\mathscr{O} = 0$. Besides, the function $f(t)$ is ($i$) operator monotone, i.e., for any density matrices $\mathscr{A}$, $\mathscr{B}$ such that $0 \leq \mathscr{A} \leq \mathscr{B}$, then $0 \leq f(\mathscr{A}) \leq f(\mathscr{B})$; ($ii$) self-inversive, $f(t) = tf(1/t)$; and ($iii$) normalized, $f(1) = 1$. Naturally, there are as many monotone metrics as there are operator monotone functions, which, according to Petz, represent a vast garden of monotone metrics~\cite{1996_LettMathPhys_38_221,*2002_JPhysA_35_929,*2013_PhysRevA_87_032324}.

As pointed out by Gibilisco and Isola~\cite{2001_InfinDimensAnalDimensQuantumProbab_44_3752}, an ordinary element of the tangent space of the density matrices manifold is given by $i[\rho,\mathcal{K}]$, where $\mathcal{K}$ is an Hermitean operator. Notably, choosing $A = B = i[\rho,\mathcal{K}]$ and taking $f(t) = (1/4){(\sqrt{t} + 1)^2}$ such that ${c_f}(x,y) = 4/{(\sqrt{x} + \sqrt{y} \, )^2}$, it follows that
\begin{align}
\label{eq:cencovmorozovfunc003}
{g_f}(i[\rho,\mathcal{K}],i[\rho,\mathcal{K}]) & = 4 \text{Tr}(i[\rho,\mathcal{K}]\, {\mathscr{M}_+^{-2}} \, 
i[\rho,\mathcal{K}]) \nonumber \\ 
& = 4 \langle {\mathscr{M}_+^{-1}} [\rho,\mathcal{K}],{\mathscr{M}_+^{-1}} 
[\rho,\mathcal{K}] \rangle  ~,
\end{align}
where ${\mathscr{M}_{\pm}} = \sqrt{\mathscr{L}} \pm \sqrt{\mathscr{R}}$ and $\langle{A},B\rangle = \text{Tr}({A^{\dagger}}B)$ is the Hilbert-Schmidt inner product. Since $[\rho,\mathcal{K}] = ({\mathscr{L}} - {\mathscr{R}})\mathcal{K} = {\mathscr{M}_+}{\mathscr{M}_-}\mathcal{K}$, we get the monotone metric
\begin{align}
\label{eq:cencovmorozovfunc004}
{g_f}(i[\rho,\mathcal{K}],i[\rho,\mathcal{K}]) & = 4 \langle {\mathscr{M}_-} \mathcal{K}, {\mathscr{M}_-} \mathcal{K}\rangle \nonumber \\ 
& = 4 \langle [ \sqrt{\rho},\mathcal{K}],[\sqrt{\rho},\mathcal{K}] \rangle \nonumber \\ 
& = - 4 \text{Tr}({[\sqrt{\rho},\mathcal{K}]^2}) \nonumber \\ 
& =  8 \mathcal{I}(\rho,\mathcal{K}) ~,
\end{align}
which, up to a constant factor, is exactly the Wigner-Yanase skew information. Recalling the multiple facets that WYSI embodies, Eq.~\eqref{eq:cencovmorozovfunc004} indicates a clearly connection between coherence measures and information geometry. It is worth mentioning that other authors also addressed the geometrical features of WYSI in a rigorous viewpoint~\cite{1999_JMathPhys_10_5702,*2002_arxiv_MRGrasselli_0212022,*2003_Jencova_2003331}. Recently, Brody~\cite{1751-8121-44-25-252002_Dorje_Brody} has demonstrated that the space of pure and mixed states is equipped with a dual metric structure which assign a clear meaning to the WYSI in the geometric realm. Besides, his approach also enabled to derive corrections to the Heisenberg uncertainty relation based on skew information.

Since the quantum state space is endowed with a metric structure, it is natural to ask about distances, curvature and other geometric properties. Particularly, the notion of distance between states has been the subject of discussions initiated decades ago under the spotlight of statistical inference~\cite{2004_PhysRevA_69_032106}. In a pioneering work, Wootters employed the statistical distance concept as a proper distinguishability measure between statistical probabilities~\cite{1981_PhysRevLett_23_357}. The geometrization of this problem emerged years later with Braunstein and Caves~\cite{1994_PhysRevLett_72_3439} who defined a Riemannian metric and its respective line element $ds$ from a suitable distinguishability quantifier between close states. Though their description was based on a physical ground, it is analogous to that one developed by Petz which relies on 
monotone metrics. Summarizing, the main message about those works lies on the close relation between state discrimination and geometric distances.

Following Petz's approach for the WYSI monotone metric, it has been proved that the distance between two density matrices on quantum state space is $\mathcal{D}(\rho,\sigma) = 2 - 2\text{Tr}(\sqrt{\rho}\sqrt{\sigma})$~\cite{2003_JMathPhys_44_3752}. This quantity is the quantum analogue of the classical Hellinger distance~\cite{2004_PhysRevA_69_032106}. Our discussion on the geometric properties of WYSI should include a few lines about geodesics -- the shortest distance between two density matrices on the quantum state space -- associated to the Wigner-Yanase monotone metric. It was shown that the corresponding geodesic distance joining the density operators $\rho$ and $\sigma$ is given by the Hellinger angle~\cite{2003_JMathPhys_44_3752,2003_arxiv_AnnaJencova_0312044}
\begin{equation}
 \label{eq:cencovmorozovfunc005}
\mathcal{L}({\rho},\sigma) = \arccos[\text{Tr}(\sqrt{\rho}\sqrt{\sigma})] ~.
\end{equation}
The quantity $\mathcal{A}(\rho,\sigma) = \text{Tr}(\sqrt{\rho}\sqrt{\sigma})$ is called {\it quantum affinity} and describes how close two states are on the quantum state space~\cite{2004_PhysRevA_69_032106}. Moreover, it is remarkable that quantum affinity is bounded from below by the Quantum Chernoff Bound (QCB)~\cite{PhysRevLett.98.160501,*2014_NewJPhys_16_073010}.


\section{Geometric lower bound on quantum coherence}
\label{sec:sectionIV}

We now provide a lower bound for the quantum coherence measure based on the skew information. Let us focus on a driven closed quantum system described initially by a mixed state $\rho_0$ which undergoes a unitary transformation ${\rho_{\varphi}} = {U_{\varphi}} \, {\rho_0}{U^{\dagger}_{\varphi}}$. Essentially, this operation encodes the parameter $\varphi$ on the input state and does not change its purity. The operator $U_{\varphi}$ characterizes a family of unitary transformations labelled by $\varphi$. Besides, it is worth to mention that $U_{\varphi}$ changes continuously with respect to this parameter. The reason for starting from a mixed state is twofold: first, the skew information is bounded by the variance when dealing with mixed states~\cite{2003_PhysRevLett_180403,2004_IEEE_50_1778}. Actually, this result was improved later by a variance lower bound which is tighter than this one based on the skew information~\cite{1751-8121-44-25-252002_Dorje_Brody}. Moreover, it also allowed to derive an entire family of higher-order corrections to the uncertainty relation supported by WYSI by exploiting its connection with the quantum analogue of the conditional variance; second, because all Petz's metrics -- particularly the Wigner-Yanase one -- becomes the well known Fubini-Study metric for pure states~\cite{Ingemar_Bengtsson_Zyczkowski}.

Considering the Wigner-Yanase metric in the quantum state space, according to Eq.~\eqref{eq:cencovmorozovfunc005} we obtain
\begin{equation}
 \label{eq:wyhel0002}
\left|\frac{d}{d\varphi}\cos[\mathcal{L}({\rho_0},{\rho_{\varphi}})]\right| =
\left|\frac{d}{d\varphi}\text{Tr}(\sqrt{\rho_0}\sqrt{\rho_{\varphi}})\right| ~.
\end{equation}
Since $U_{\varphi}$ is a unitary operator, it is possible to write $\sqrt{\rho_{\varphi}} = {U_{\varphi}}\sqrt{\rho_0}{U_{\varphi}^{\dagger}}$ (see Appendix~\ref{sec:appendixA}) which implicates the von Neumann equation
\begin{equation}
 \label{eq:wyhel0004}
\frac{d\sqrt{\rho_{\varphi}}}{d\varphi} = -\frac{i}{\hbar}[{K_{\varphi}},\sqrt{\rho_{\varphi}}] ~,
\end{equation}
where we used that $({dU_{\varphi}}/{d\varphi}){U_{\varphi}^{\dagger}} = -{U_{\varphi}}({dU_{\varphi}^{\dagger}}/{d\varphi})$ and defined the Hermitian operator
\begin{equation}
 \label{eq:wyhel0005}
{K_{\varphi}} = -i{\hbar}{U_{\varphi}}\frac{dU_{\varphi}^{\dagger}}{d\varphi} ~.
\end{equation}
In general, the operator $K_{\varphi}$ depends on the parameter $\varphi$. However, it is worth to notice that as a special case, when the observable ${K_{\varphi}}$ is independent of $\varphi$, i.e., ${K_{\varphi}} = K$, thus the unitary evolution is given by ${U_{\varphi}} = {e^{-i\varphi K}}$. 

Returning to the general case and substituting Eq.~\eqref{eq:wyhel0004} into Eq.~\eqref{eq:wyhel0002}, we have
\begin{equation}
 \label{eq:wyhel000502}
\left|\frac{d}{d\varphi}\cos[\mathcal{L}({\rho_0},{\rho_{\varphi}})]\right| =
\frac{1}{\hbar}\left|\text{Tr}(\sqrt{\rho_0}[{K_{\varphi}},\sqrt{\rho_{\varphi}}])\right| ~.
\end{equation}
Equation~\eqref{eq:wyhel000502} is the starting point for establishing the lower bound on the skew information. Actually, this goal is reached by noting that
\begin{equation}
 \label{eq:wyhel0006}
\left|\text{Tr}(\sqrt{\rho_0}[{K_{\varphi}},\sqrt{\rho_{\varphi}}])\right| \leq \\ 
{\|\sqrt{\rho_0}\|_2}{\|[{K_{\varphi}},\sqrt{\rho_{\varphi}}]\|_2} ~,
\end{equation}
where we used the Cauchy-Schwarz inequality $|\text{Tr}(AB)| \leq {\| A \|_2}{\| B \|_2}$, with ${\| A \|_2} = \sqrt{\text{Tr}({A^{\dagger}}A)}$ being the Schatten $2-$norm (also known as Hilbert Schmidt or Fr\"{o}benius norm)~\cite{Bathia_Rajendra}. Combining Eq.~\eqref{eq:wyhel0006} with ${\|\sqrt{\rho_0} \|_2} = 1$ and substituting the result into Eq.~\eqref{eq:wyhel000502}, we obtain
\begin{equation}
 \label{eq:wyhel0007}
\left|\frac{d}{d\varphi}\cos[\mathcal{L}({\rho_0},{\rho_{\varphi}})]\right| \leq 
\frac{1}{\hbar}{\|[{K_{\varphi}},\sqrt{\rho_{\varphi}}]\|_2} ~.
\end{equation}
On the other hand, note that
\begin{equation}
 \label{eq:wyhel0008}
{\|[{K_{\varphi}},\sqrt{\rho_{\varphi}}]\|_2}  = \sqrt{-\text{Tr}({[\sqrt{\rho_{\varphi}},{K_{\varphi}}]^2})} 
= \sqrt{2\mathcal{I}({\rho_{\varphi}},{K_{\varphi}})} ~,
\end{equation}
where $\mathcal{I}({\rho_{\varphi}},{K_{\varphi}}) = -(1/2)\text{Tr}([\sqrt{\rho_{\varphi}},{K_{\varphi}}]^2)$ is the Wigner-Yanase skew information between the evolved state $\rho_{\varphi}$ and the observable $K_{\varphi}$ that generates the dynamics. Therefore, substituting Eq.~\eqref{eq:wyhel0008} into Eq.~\eqref{eq:wyhel0007} we obtain a lower bound in terms of WYSI and Hellinger angle as follows
\begin{equation}
 \label{eq:wyhel00010}
\left|\frac{d}{d\varphi}\cos[\mathcal{L}({\rho_0},{\rho_{\varphi}})]\right| \leq
\frac{\sqrt{2}}{\hbar}\sqrt{\mathcal{I}({\rho_{\varphi}},{K_{\varphi}})}  ~.
\end{equation}

Eq.~\eqref{eq:wyhel00010} is the main result of this paper. It is important to highlight that it encompasses any class of continuous unitary transformations $U_{\varphi}$ indexed by the parameter $\varphi$, as well as initial and evolved mixed states. As a particular case, recalling that a unitary evolution does not change the purity of a quantum state, if $\rho_0$ is pure, then $\rho_{\varphi}$ will also be, and the skew information reduces to the variance of $K_{\varphi}$, i.e., $\mathcal{I}({\rho_{\varphi}},{K_{\varphi}}) = {(\Delta{K_{\varphi}})^2} = \langle{K_{\varphi}^2}\rangle - {\langle{K_{\varphi}}\rangle^2}$. In this regime, the lower bound becomes
\begin{equation}
 \label{eq:wyhel0001001}
 {\Delta{K_{\varphi}}} \geq \frac{\hbar}{\sqrt{2}}\left|\frac{d}{d\varphi}f(\varphi)\right| ~,
\end{equation}
where $f(\varphi) = \text{Tr}[{\rho_0}{\rho_{\varphi}}]/\text{Tr}{\rho_0^2}$ defines the relative purity, which played a special role for the investigation of quantum speed limits under the closed dynamics~\cite{2009_PhysRevLett_103_160502}.


\section{Example}
\label{sec:sectionV}

To illustrate the use of the bound indicated in Eq.~\eqref{eq:wyhel00010}, we now consider the single qubit case. Let ${\rho_{0}} = ({1}/{2})(I + {\vec{r}_0}\cdot\vec{\sigma})$ be the initial state ($I$ denotes the $2\times2$ identity matrix, ${\vec{r}_0}$ is a 3-dimensional vector with ${|{\vec{r}_0}|^2} = {r_0^2}< 1$ and $\vec{\sigma} = \{{\sigma_1},{\sigma_2},{\sigma_3}\}$ is the vector of the Pauli matrices). The dynamics is governed by the self-commuting local observable ${K_{\varphi}} = \varpi(\alpha{I} + {\hat{n}_{\varphi}}\cdot\vec{\sigma})$, i.e., $[{K_{\varphi}},{K_{\varphi'}}] = 0$ for all $\varphi$ and $\varphi'$, where $\varpi$ and $\alpha$ are positive constants and ${\hat{n}_{\varphi}}$ is an unit vector, $|{\hat{n}_{\varphi}}| = 1$. The system evolves under a general unitary operator ${U_{\varphi}}$ given by
\begin{align}
 \label{eq:example002}
{U_{\varphi}} & = \exp\left[-\frac{i}{\hbar}{\int_0^{\varphi}}d{\varphi'}{K_{\varphi'}}\right] \nonumber \\
& = {e^{-i\delta\alpha}}[{I}\cos\gamma - i({\hat{\Sigma}_{\varphi}}\cdot\vec{\sigma})
\sin\gamma] ~,
\end{align}
where $\gamma = {{\varpi}}\varphi|{\vec{\Sigma}_{\varphi}}|/{\hbar}$ and $\delta = {{\varpi}}\varphi/{\hbar}$ are dimensionless constants and also
\begin{equation}
 \label{eq:example003}
{\vec{\Sigma}_{\varphi}} := \frac{1}{\varphi}{\int_0^{\varphi}}d{\varphi'}{\hat{n}_{\varphi'}} ~.
\end{equation}

Essentially, the initial state $\rho_0$ undergoes the unitary transformation ${\rho_{\varphi}} = {U_{\varphi}} \, {\rho_0}{U^{\dagger}_{\varphi}}$ which encodes the parameter $\varphi$. It is possible to verify that the final state can be written as ${\rho_{\varphi}} = ({1}/{2})(I + {\vec{r}_{\varphi}}\cdot\vec{\sigma})$, where (see Appendix~\ref{sec:appendixB})
\begin{equation}
 \label{eq:example006}
{\vec{r}_{\varphi}} = \cos(2\gamma){\vec{r}_0} + [1 - \cos(2\gamma)]({\hat{\Sigma}_{\varphi}}
\cdot{\vec{r}_0}){\hat{\Sigma}_{\varphi}} + \sin(2\gamma)({\hat{\Sigma}_{\varphi}}\times{\vec{r}_0}) ~.
\end{equation}
The vector ${\vec{r}_{\varphi}}$ keeps whole information about the parameter $\varphi$ and has the same magnitude as the initial vector $\vec{r}_0$, i.e., $|{\vec{r}_{\varphi}}| = |{\vec{r}_0}| = {r_0}$. Particularly, as a special case, if ${\hat{n}_{\varphi}}$ is independent of the parameter $\varphi$, i.e., ${\hat{n}_{\varphi}} = \hat{n}$, then Eq.~\eqref{eq:example003} implies that ${\hat{\Sigma}_{\varphi}} = \hat{n}$ and, consequently, $\gamma = \delta = {{\varpi}}\varphi/{\hbar}$. 

In order to calculate the Hellinger angle we need to determine the trace of the product of operators $\sqrt{\rho_0}$ and $\sqrt{\rho_{\varphi}}$. The analytical expressions for the square root of a single qubit state can be found in Appendix~\ref{sec:appendixC}. Since the modulus of Bloch sphere radius 
remains constant under the unitary transformation, it is possible to verify that the cosine of the Hellinger angle becomes
\begin{equation}
 \label{eq:example0015}
\cos[\mathcal{L}({\rho_0},{\rho_{\varphi}})] = \frac{1}{2}[{\xi_+} + {\xi_-}({\hat{r}_{\varphi}}\cdot{\hat{r}_0})] ~,
\end{equation}
where ${\xi_{\pm}} = 1 \pm \sqrt{1 - {r_0^2}}$ is independent of the parameter $\varphi$. From this result is straightforward to check 
that $d\cos[\mathcal{L}({\rho_0},{\rho_{\varphi}})]/d\varphi = ({\xi_-}/2)[({d{{\hat{r}}_{\varphi}}/d\varphi})\cdot{\hat{r}_0}]$. Similarly, the Wigner-Yanase 
skew information is given by
\begin{equation}
 \label{eq:example0016}
\mathcal{I}({\rho_{\varphi}},{K_{\varphi}}) = {\varpi^2}{\xi_-}{|{\hat{r}_{\varphi}}\times{\hat{n}_{\varphi}}|^2} ~.
\end{equation}
Substituting the derivative of Eq.~\eqref{eq:example0015} on the parameter $\varphi$ and Eq.~\eqref{eq:example0016} into Eq.~\eqref{eq:wyhel00010}, we 
finally obtain the bound $\sqrt{\xi_-}|({d{{\hat{r}}_{\varphi}}/d\varphi})\cdot{\hat{r}_0}| \leq 2\sqrt{2}(\varpi/\hbar)|{\hat{r}_{\varphi}}\times{\hat{n}_{\varphi}}|$. 
To clarify, choosing the probe state ${\rho_0} = (1 - {r_0})I + {r_0}|\psi\rangle\langle{\psi}|$, $|\psi\rangle = (1/\sqrt{2})(|0\rangle + |1\rangle)$, and $\hat{n} = 
(0,0,1)$, which corresponds to take ${\vec{r}_0} = {r_0}(\cos\phi,\sin\phi,0)$ ($0 < {r_0} < 1$ and $0 \leq \phi \leq 2\pi$) and ${K_{\varphi}} = \varpi(\alpha I + {\sigma_z})$, the cosine of the Hellinger angle is $\cos[\mathcal{L}({\rho_0},{\rho_{\varphi}})] = (1/2)[{\xi_+} + {\xi_-}\cos(2\varphi)]$ and the WYSI gives $\mathcal{I}({\rho_{\varphi}},{K_{\varphi}}) = {\varpi^2}{\xi_-}{\sin^2}\phi$. Combining both results we obtain the bound $\sqrt{\xi_-}|\sin(2\varphi)| \leq \sqrt{2}(\varpi/\hbar)
|\sin\phi|$. It is interesting to note that, while $\mathcal{L}({\rho_0},{\rho_{\varphi}})$ is a function of the parameter $\varphi$, WYSI is independent of this phase 
and describes a constant of motion during the unitary evolution. It is worth mentioning that although we focused attention in the single qubit case, our calculations 
can be extended to a system of $N$ qubits.


\section{Conclusions}
\label{sec:sectionVI}

In this article we established a geometric lower bound for a proper quantum coherence measure based on Wigner-Yanase skew information. This information-theoretic quantity is regarded as a particular extension of Fisher information and can be seen as a monotone Riemannian metric due to the Petz's theorem~\cite{1996_LinAlgApl_244_81}. Moreover, its related geodesic distance on quantum space state is given by the Hellinger angle~\cite{2003_JMathPhys_44_3752}. In opposition to many other distance measures such as Bures angle or even relative entropy, Hellinger angle is advantageous quantity because is technically easier to calculate and more intuitive to obtain from its classical statistical analogous. Despite those motivational issues, it has received little attention beyond that devoted to the exploration of its useful algebraic properties to the information theory. It is important to emphasize that our result shows that, since geodesic distance quantifies the discrimination of two density operators in the context of quantum statistical estimation theory~\cite{2004_PhysRevA_69_032106}, skew information is bounded from below by the rate of change of distinguishability between two states on quantum state space.

Our result opens a wide range of possible physical interpretations. First, inequality Eq.~\eqref{eq:wyhel00010} suggests a route for better understanding the phase estimation paradigm in quantum metrology~\cite{PhysRevLett.96_010401,*2009_Int_J_Quantum_Inform_7_125_Paris,*NatPhoton_5_222_GSLM,*PhysRevLett_112.210401}. In fact, it can provide a precision bound for an unknown parameter $\varphi$ encoded by the unitary transformation in the initial state. Therefore, the bound essentially depends on the derivative of the Hellinger angle with respect to this parameter.

In particular, choosing $\varphi = \tau$, where $\tau$ is time, it can be shown that our inequality gives rise to a new quantum speed limit~\cite{Ourspeedlimit}. In contrast with the original one proposed by Mandelstamm-Tamm~\cite{1945_JPhysURSS_9_249}, and later generalizations for driven closed systems~\cite{1993_PhysRevLett_70_3365,*2003_PhysRevA_67_052109,*2003_PhysRevA_82_022107,*2013_JMathPhys_46_335302,*2013_PhysRevLett_111_010402}, this speed limit depends on WYSI and the Hellinger angle rather than the Bures angle or the variance of the hamiltonian.

Besides, it seems possible to attribute a thermodynamic meaning for this bound by investigating the connection between nonequilibrium entropy production~\cite{2011_PhysRevLett_107_140404} and the thermodynamic length~\cite{2007_PhysRevLett_100602,*2013_PhysRevE_87_022143} involving quantum protocols at finite temperature. This could provide a thermodynamic interpretation for the existence of the quantum speed limit. 

Finally, in a future work it will be crucial to investigate the eventual relation between geometric bounds and the universality class of Petz metrics which fulfils the requirements for a quantum coherence measure. Moreover, to enlarge the present analysis, take into account the open quantum dynamics would be essential not only for the foundations of quantum information theory but also for realizing quantum technology in a noisy scenario. From the experimental point of view, by extending our conclusions to $N$ quibt systems, the bound in Eq. (\ref{eq:wyhel00010}) could be experimentally investigated through a measurement scheme based on two-point correlation functions~\cite{2013_arXiv_13124240v1_BuscemiDallArnoOzawaVedral}.


\section*{Acknowledgments}

We are grateful to M. Cianciaruso, Dr. G. Adesso, and Prof. D. C. Brody for fruitful correspondence. The authors would like to thank the financial support from the Brazilian funding agencies CNPq and CAPES. This work was also supported by the Brazilian National Institute for Science and Technology of Quantum Information (INCT-IQ).

\setcounter{equation}{0}
\setcounter{figure}{0}
\setcounter{table}{0}
\setcounter{section}{0}
\makeatletter
\renewcommand\thesection{\Alph{section}}
\def\theequation{\thesection.\arabic{equation}}
\renewcommand{\thefigure}{\arabic{figure}}
\renewcommand{\bibnumfmt}[1]{[#1]}
\renewcommand{\citenumfont}[1]{#1}

\section*{Appendix}

\section{Matrix powers}
\label{sec:appendixA}

In this section we will demonstrate that the identity ${(V \Lambda {V^{\dagger}})^s} = V{\Lambda^s}{V^{\dagger}}$ still holds for $0 < s < 1$, where $\Lambda$ is a positive matrix $(\Lambda > 0)$ and $V$ is an unitary operator, ${V^{\dagger}} = {V^{-1}}$. In order to reach the main goal, let us consider a monotone function $f(a) = {a^s}$ for $a > 0$. It can be demonstrated that $f(a)$ has the following integral representation~\cite{Bathia_Rajendra}
\begin{equation}
 \label{eq:matrixpower001}
{a^s} = \frac{\sin(\pi s)}{\pi s}{\int_0^{\infty}} \frac{a}{a + x}d\mu(x) ~,
\end{equation} 
where $\mu (x) = {x^s}$, $d\mu(x) = s{x^{s-1}}dx$, is a positive measure on $(0,\infty)$. This relation can be extended to 
the positive and non-singular operator $\Lambda$ as follows~\cite{LinMing_JPhysAMathTheor_47_055301} 
\begin{equation}
 \label{eq:matrixpower002}
{\Lambda^s} = \frac{\sin(\pi s)}{\pi s}
{\int_0^{\infty}}{\Lambda}{(\Lambda + xI)^{-1}} d\mu(x) ~.
\end{equation} 
Actually, the last condition can be relaxed if $\Lambda$ is full rank or if we assume that the inverse operation is taken 
on its suport, i.e., the vector subspace spanned by the eigenstates with non-zero eigenvalues~\cite{Campbell}. Considering 
the transformation $V\Lambda{V^{\dagger}}$, follows
\begin{align}
 \label{eq:matrixpower0031}
{(V\Lambda{V^{\dagger}} + xI)^{-1}} & = {[V(\Lambda + xI){V^{\dagger}}]^{-1}} \nonumber \\ 
& =  V{(\Lambda + xI)^{-1}}{V^{\dagger}} ~.
\end{align}
Therefore, we can prove our main goal combining the previous equality with the integral representation indicated 
in Eq.~\eqref{eq:matrixpower002}, i.e.,
\begin{align}
 \label{eq:matrixpower004}
{(V\Lambda{V^{\dagger}})^s} & = \frac{\sin(\pi s)}{\pi s}{\int_0^{\infty}}V
\Lambda{V^{\dagger}}{(V\Lambda{V^{\dagger}} + xI)^{-1}}d\mu(x) \nonumber \\ 
& = V\frac{\sin(\pi s)}{\pi s}{\int_0^{\infty}}\Lambda{(\Lambda + xI)^{-1}}d\mu(x){V^{\dagger}} \nonumber \\ 
& = V{\Lambda^s}{V^{\dagger}} ~.
\end{align}
Particularly, given the evolved state ${\rho_{\varphi}} =  {U_{\varphi}}\, {\rho_0}{U_{\varphi}^{\dagger}}$, choosing $\Lambda = {\rho_0}$ as the initial mixed state 
and the unitary operator $V = {U_{\varphi}}$, for $s = 1/2$ Eq.~\eqref{eq:matrixpower004} allows to demonstrate the relation $\sqrt{\rho_{\varphi}} = {U_{\varphi}}\sqrt{\rho_0}{U_{\varphi}^{\dagger}}$.

\section{Unitary evolution}
\label{sec:appendixB}

In this section we describe the calculation of the evolved state ${\rho_{\varphi}}$ in the single 
qubit context. Let us assume that the quantum system dynamics is governed by the local 
observable ${K_{\varphi}} = \varpi(\alpha{I} + {\hat{n}_{\varphi}}\cdot\vec{\sigma})$, where 
$\varpi$ and $\alpha$ are positive constants and ${\hat{n}_{\varphi}}$ is an unit vector, i.e., 
$|{\hat{n}_{\varphi}}| = 1$. By hypothesis, this observable is self-commuting, i.e., 
$[{K_{\varphi}},{K_{\varphi'}}] = 0$ for all $\varphi$ and $\varphi'$. The system evolves 
under a general unitary operator ${U_{\varphi}}$ given by
\begin{align}
 \label{eq:genexample002}
{U_{\varphi}} & = \exp\left[-\frac{i}{\hbar}{\int_0^{\, \varphi}}d{\varphi'}{K_{\varphi'}}\right] \nonumber \\ 
& = {e^{-i\delta\alpha}}\exp[-i\gamma({\hat{\Sigma}_{\varphi}}\cdot\vec{\sigma})] \nonumber \\
& = {e^{-i\delta\alpha}}[{I}\cos\gamma - i({\hat{\Sigma}_{\varphi}}\cdot\vec{\sigma})
\sin\gamma] ~,
\end{align}
where $\gamma = {{\varpi}}|{\vec{\Sigma}_{\varphi}}|\varphi/{\hbar}$ and 
$\delta = {{\varpi}}\varphi/{\hbar}$ are dimensionless constants 
and also
\begin{equation}
 \label{eq:genexample003}
{\vec{\Sigma}_{\varphi}} := \frac{1}{\varphi}{\int_0^{\varphi}}d{\varphi'}{\hat{n}_{\varphi'}} ~.
\end{equation}
In particular, if ${\hat{n}_{\varphi}}$ is independent of the parameter $\varphi$, i.e., ${\hat{n}_{\varphi}} = 
\hat{n}$, then ${\hat{\Sigma}_{\varphi}} = \hat{n}$. Returning to the general case, let be an initial single 
qubit mixed state ${\rho_0} = ({1}/{2})(I + {\vec{r}_0}\cdot\vec{\sigma})$, where $I$ denotes the $2\times2$ 
identity matrix, ${|{\vec{r}_0}|^2} = {r_0^2} < 1$ and $\vec{\sigma} = \{{\sigma_1},{\sigma_2},{\sigma_3}\}$ is 
a vector of the Pauli matrices. The probe state $\rho_0$ undergoes the unitary transformation ${\rho_{\varphi}} 
= {U_{\varphi}}{\rho_0}{U^{\dagger}_{\varphi}}$ and can be written as
\begin{multline}
 \label{eq:genexample005}
{\rho_{\varphi}} = \frac{1}{2}\{I + ({\vec{r}_0}\cdot\vec{\sigma}){\cos^2}\gamma + 
i[{\vec{r}_0}\cdot\vec{\sigma},{\hat{\Sigma}_{\varphi}}\cdot\vec{\sigma}]\sin\gamma\cos\gamma + \\ +
({\hat{\Sigma}_{\varphi}}\cdot\vec{\sigma})({\vec{r}_0}\cdot\vec{\sigma})(
{\hat{\Sigma}_{\varphi}}\cdot\vec{\sigma}){\sin^2}\gamma\} ~.
\end{multline}
Exploring the algebraic properties of Pauli matrices is possible to check that 
$(\vec{a}\cdot\vec{\sigma})(\vec{b}\cdot\vec{\sigma}) = (\vec{a}\cdot\vec{b})I + 
i(\vec{a}\times\vec{b})\cdot\vec{\sigma}$. Combining this relation with the
vector identities $\vec{a}\cdot(\vec{a}\times\vec{b}) = 0$ and $\vec{a}\times(
\vec{b}\times\vec{c}) = (\vec{a}\cdot\vec{c})\vec{b} - (\vec{a}\cdot\vec{b})
\vec{c}$, we obtain
\begin{equation}
  \label{eq:genexample00601}
[{\vec{r}_0}\cdot\vec{\sigma},{\hat{\Sigma}_{\varphi}}\cdot\vec{\sigma}] = - 2i({\hat{\Sigma}_{\varphi}}
\times{\vec{r}_0})\cdot\vec{\sigma}
\end{equation}
and
\begin{equation}
  \label{eq:genexample00602}
({\hat{\Sigma}_{\varphi}}\cdot\vec{\sigma})({\vec{r}_0}\cdot\vec{\sigma})(
{\hat{\Sigma}_{\varphi}}\cdot\vec{\sigma}) = [2({\hat{\Sigma}_{\varphi}}\cdot{\vec{r}_0})
{\hat{\Sigma}_{\varphi}} - {\vec{r}_0}]\cdot\vec{\sigma} ~.
\end{equation}
Substituting Eq.~\eqref{eq:genexample00601}--\eqref{eq:genexample00602} into Eq.~\eqref{eq:genexample005} 
and performing the calculations, it is possible to verify that the evolved state becomes ${\rho_{\varphi}} 
= ({1}/{2})(I + {\vec{r}_{\varphi}}\cdot\vec{\sigma})$, with
\begin{equation}
 \label{eq:genexample008}
{\vec{r}_{\varphi}} = \cos(2\gamma){\vec{r}_0} + [1 - \cos(2\gamma)]({\hat{\Sigma}_{\varphi}}
\cdot{\vec{r}_0}){\hat{\Sigma}_{\varphi}} + \sin(2\gamma)({\hat{\Sigma}_{\varphi}}\times{\vec{r}_0}) ~.
\end{equation}
It is worth to emphasize that both vectors $\vec{r}_{\varphi}$ and $\vec{r}_0$ has the same absolute 
value, ${|{\vec{r}_{\varphi}}|^2} = {|{\vec{r}_0}|^2} = {r_0^2}$. In other words, the unitary transformation 
$U_{\varphi}$ does not change the modulus of Bloch sphere radius during the dynamics. 

Now we will provide another proof for Eq.~\eqref{eq:genexample008} which is based on the 
{\it Rodrigues' rotation formula}. Summarizing, through this approach the vector $\vec{r}_{\varphi}$ is 
completely determined by the action of a rotation matrix on the initial vector $\vec{r}_0$. In order to 
understand this property, let be the $j-${\it th} component ${({\vec{r}_{\varphi}})_j} = 
{({S_{\varphi}})_{jl}}{({\vec{r}_0})_l}$, with
\begin{equation}
 \label{eq:genexample009}
{({S_{\varphi}})_{jl}} = \cos(2\gamma){\delta_{jl}} + [1 - \cos(2\gamma)]{({\hat{\Sigma}_{\varphi}})_{j}}
{({\hat{\Sigma}_{\varphi}})_{l}} + \sin(2\gamma){({\Lambda_{\varphi}})_{jl}}
\end{equation}
and ${({\Lambda_{\varphi}})_{jl}} := {\epsilon_{jkl}}{({\hat{\Sigma}_{\varphi}})_{k}}$. The matrix element 
${({\Lambda_{\varphi}})_{jl}}$ satisfy the identity
\begin{align}
 \label{eq:genexample010}
{({\Lambda_{\varphi}})_{js}}{({\Lambda_{\varphi}})_{sl}} & = {\epsilon_{sjk}}{\epsilon_{s{\mu}l}}{({\hat{\Sigma}_{\varphi}})_{k}}
{({\hat{\Sigma}_{\varphi}})_{\mu}} \nonumber \\
& = ({\delta_{j\mu}}{\delta_{kl}} - {\delta_{jl}}{\delta_{k\mu}}){({\hat{\Sigma}_{\varphi}})_{k}}
{({\hat{\Sigma}_{\varphi}})_{\mu}} \nonumber \\ 
& = {({\hat{\Sigma}_{\varphi}})_{j}}{({\hat{\Sigma}_{\varphi}})_{l}} - {\delta_{jl}} ~,
\end{align}
where we used the Einstein summation convention and the property ${({\hat{\Sigma}_{\varphi}})_{k}}{({\hat{\Sigma}_{\varphi}})_{k}} = 
{|{\hat{\Sigma}_{\varphi}}|^2} = 1$. From this expression we have ${({\hat{\Sigma}_{\varphi}})_{j}}{({\hat{\Sigma}_{\varphi}})_{l}} = 
{\delta_{jl}} + {\Lambda_{js}}(\varphi){\Lambda_{sl}}(\varphi)$ and therefore
\begin{equation}
 \label{eq:genexample011}
{({S_{\varphi}})_{jl}} = {\delta_{jl}} + [1 - \cos(2\gamma)]{({\Lambda_{\varphi}})_{js}}{({\Lambda_{\varphi}})_{sl}} + 
\sin(2\gamma){({\Lambda_{\varphi}})_{jl}} ~.
\end{equation}
The matrix $\Lambda_{\varphi}$ is called {\it skew tri-idempotent} because fulfils ${\Lambda_{\varphi}^3} = 
- {\Lambda_{\varphi}}$. This property can be verified starting from the triple product
\begin{align}
 \label{eq:genexample012}
{({\Lambda_{\varphi}})_{js}}{({\Lambda_{\varphi}})_{s\mu}}{({\Lambda_{\varphi}})_{\mu l}} & =
{\epsilon_{{\mu}{\alpha}l}}{({\hat{\Sigma}_{\varphi}})_{\mu}}{({\hat{\Sigma}_{\varphi}})_{\alpha}}{({\hat{\Sigma}_{\varphi}})_{j}} - 
{\delta_{j\mu}}{\epsilon_{{\mu}{\alpha}l}}{({\hat{\Sigma}_{\varphi}})_{\alpha}} \nonumber \\
& = [{\epsilon_{l{\mu}{\alpha}}}{({\hat{\Sigma}_{\varphi}})_{\mu}}{({\hat{\Sigma}_{\varphi}})_{\alpha}}]{({\hat{\Sigma}_{\varphi}})_{j}} - 
{\epsilon_{j{\alpha}l}}{({\hat{\Sigma}_{\varphi}})_{\alpha}} \nonumber \\
& = - {({\Lambda_{\varphi}})_{jl}} ~.
\end{align}
Note that the last equality in the expression above was obtained by using the identity $\hat{l}\cdot({\hat{\Sigma}_{\varphi}}\times
{\hat{\Sigma}_{\varphi}}) = {\epsilon_{l{\mu}{\alpha}}}{({\hat{\Sigma}_{\varphi}})_{\mu}}{({\hat{\Sigma}_{\varphi}})_{\alpha}} = 
0$. From the result obtained in Eq.~\eqref{eq:genexample012} the matrix $S_{\varphi}$ can be written as
\begin{align}
 \label{eq:genexample013}
{S_{\varphi}} & = I + [1 - \cos(2\gamma)]{\Lambda_{\varphi}^2} + \sin(2\gamma){\Lambda_{\varphi}} \nonumber \\
& =  {e^{2\gamma{\Lambda_{\varphi}}}} ~.
\end{align}
From this relation is possible to identify the explicity form of matrix ${\Lambda_{\varphi}}$. First, this matrix has all 
diagonal elements equal to zero, i.e., ${({\Lambda_{\varphi}})_{jj}} = {\epsilon_{jkj}}{({\hat{\Sigma}_{\varphi}})_{k}} = 0$. 
Second, the matrix $\Lambda_{\varphi}$ is anti-symmetric because ${({\Lambda_{\varphi}})_{lj}} = {\epsilon_{lkj}}
{({\hat{\Sigma}_{\varphi}})_{k}} = -{\epsilon_{jkl}}{({\hat{\Sigma}_{\varphi}})_{k}} = -{({\Lambda_{\varphi}})_{jl}}$. On the other 
hand, given that ${({\Lambda_{\varphi}})_{23}} = {\epsilon_{213}}{({\hat{\Sigma}_{\varphi}})_{1}} = -{({\hat{\Sigma}_{\varphi}})_{1}}$, 
${({\Lambda_{\varphi}})_{13}} = {\epsilon_{123}}{({\hat{\Sigma}_{\varphi}})_{2}} = -{({\hat{\Sigma}_{\varphi}})_{2}}$ and 
${({\Lambda_{\varphi}})_{12}} = {\epsilon_{132}}{({\hat{\Sigma}_{\varphi}})_{3}} = -{({\hat{\Sigma}_{\varphi}})_{3}}$, is 
immediate to write $\Lambda_{\varphi}$ as
\begin{equation}
 \label{eq:genexample015}
{\Lambda_{\varphi}} = -i{\hat{\Sigma}_{\varphi}}\cdot\vec{J} ~,
\end{equation}
where $\vec{J} = \{{J_1},{J_2},{J_3}\}$ is a vector whose components are given by the generators 
of the adjoint representation ($3-$dimensional) of SU$(2)$ algebra,
\begin{equation}
 \label{eq:genexample016}
{J_1} = \left[\begin{matrix} 0 & 0 & 0 \\ 0 & 0 & -i \\ 0 & i & 0 \end{matrix}\right] ~,\quad
{J_2} = \left[\begin{matrix} 0 & 0 & i \\ 0 & 0 & 0 \\ -i & 0 & 0 \end{matrix}\right] ~,\quad
{J_3} = \left[\begin{matrix} 0 & -i & 0 \\ i & 0 & 0 \\ 0 & 0 & 0 \end{matrix}\right] ~.
\end{equation}
Finally, substituting Eq.~\eqref{eq:genexample015} into Eq.~\eqref{eq:genexample013}, the vector 
$\vec{r}_{\varphi}$ is written as follows
\begin{equation}
 \label{eq:genexample018}
{\vec{r}_{\varphi}} = {e^{-i2\gamma{\hat{\Sigma}_{\varphi}}\cdot\vec{J}}}{\vec{r}_0} ~.
\end{equation}

\section{Hellinger angle and WYSI: single qubit case}
\label{sec:appendixC}

In this section we provide an explicit calculation of the Hellinger angle and Wigner-Yanase skew information 
for a mixed single qubit state. To achieve these results we will obtain analytical expressions of the inverse matrix, 
determinant and square root for that state. Consider a hermitian {\it contractive} operator $\varPi$, i.e., $\|\varPi\| 
\leq 1$, where $\|\ldots\|$ defines the operator (or bound) norm. In this case, the positive operator $I + \varPi$ is 
invertible and its inverse
\begin{align}
 \label{eq:wyhel00020}
{(I + \varPi)^{-1}} & = I - \varPi + {\varPi^2} - {\varPi^3} + \ldots \nonumber \\
& = (I - \varPi)(I + {\varPi^2} + {\varPi^4} + \ldots) \nonumber \\ 
& = (I - \varPi){(I - {\varPi^2})^{-1}}
\end{align}
defines a convergent {\it Neumann series}~\cite{Bathia_Rajendra}. Let us consider now a single qubit 
mixed state ${\rho_{\mu}} = (1/2)(I + {\vec{r}_{\mu}}\cdot\vec{\sigma})$ with $\mu \in \{0,\varphi\}$ and 
choose $\varPi = {\vec{r}_{\mu}}\cdot\vec{\sigma}$. Here $I$ denotes the $2\times2$ identity matrix, 
$\vec{\sigma} = ({\sigma_1},{\sigma_2},{\sigma_3})$ and ${\vec{r}_{\mu}}$ is a 3-dimensional vector which fulfils 
${|{\vec{r}_{\mu}}|^2} < 1$. By using the vector identity $(\vec{a}\cdot\vec{\sigma})(\vec{b}\cdot\vec{\sigma}) = 
(\vec{a}\cdot\vec{b})I + i(\vec{a}\times\vec{b})\cdot\vec{\sigma}$, is straightforward to verify ${\varPi^2} = 
{({\vec{r}_{\mu}}\cdot\vec{\sigma})^2} = {|{\vec{r}_{\mu}}|^2}I$ and $\text{Tr}(\varPi) = 0$. Given that $\|\varPi\| = 
\||\varPi|\|$, where $|\varPi| = \sqrt{{\varPi^{\dagger}}\varPi} = |{\vec{r}_{\mu}}|I$, in our case follows $\|\varPi\| = 
|{\vec{r}_{\mu}}|\|I\| = |{\vec{r}_{\mu}}| \leq 1$ and according to Eq.~\eqref{eq:wyhel00020} we obtain
\begin{equation}
 \label{eq:wyhel00021}
{\rho_{\mu}^{-1}} = 2{(I + {\vec{r}_{\mu}}\cdot\vec{\sigma})^{-1}} = 
\frac{2}{1 - {|{\vec{r}_{\mu}}|^2}}({I - {\vec{r}_{\mu}}\cdot\vec{\sigma}}) ~.
\end{equation}
Note that the previous result is singular if the state is a pure one. Actually, in this case the inverse operation 
requires another approach known as {\it generalized inverse} or {\it Moore-Penrose inverse}~\cite{Penrose}. 
Returning to the mixed case, it is a simple task to recognize $1 - {|{\vec{r}_{\mu}}|^2}$ as the 
determinant of the state $\rho_{\mu}$ starting from the identity
\begin{equation}
 \label{eq:wyhel00022}
\det(I + \varPi) = {e^{\text{Tr}[\ln(I + \varPi)]}} ~.
\end{equation}
In fact, since $\|\varPi\| \leq 1$ and taking the Taylor series expansion $\ln(1 + x) = -{\sum_{k=1}^{\infty}}{(-x)^k}/k$ 
for  $|x| < 1$, follows
\begin{align}
 \label{eq:wyhel00023}
\text{Tr}[\ln(I + \varPi)] & = -{\sum_{k = 1}^{\infty}}\frac{(-1)^{k}}{k}\text{Tr}(\varPi^{k}) \nonumber \\ 
& = -{\sum_{k=1}^{\infty}}\frac{{|{\vec{r}_{\mu}}|^{2k}}}{k} \nonumber \\ 
& = \ln(1 - {|{\vec{r}_{\mu}}|^2}) ~,
\end{align}
where we used $\text{Tr}(\varPi^{2k+1}) = 0$ and $\text{Tr}(\varPi^{2k}) = 2{|{\vec{r}_{\mu}}|^{2k}}$ and 
collected separately even and odd contributions in the infinite sum. Therefore, we get
\begin{equation}
 \label{eq:wyhel00024}
\det(I + {\vec{r}_{\mu}}\cdot\vec{\sigma}) = 1 - {|{\vec{r}_{\mu}}|^2} ~.
\end{equation}

In order to calculate the square root of the density operator $\rho_{\mu}$, it is convenient to 
remember the integral representation presented in Eq.~\eqref{eq:matrixpower002} choosing now 
$\Lambda = {\rho_{\mu}}$ and $s = 1/2$, i.e.,
\begin{equation}
 \label{eq:wyhel00025}
\sqrt{\rho_{\mu}} = \frac{1}{\pi}{\int_0^{\infty}}\frac{dx}{\sqrt{x}}{\rho_{\mu}}
{({\rho_{\mu}} + xI)^{-1}} ~.
\end{equation}
According to Eq.~\eqref{eq:wyhel00021} it can be verified that
\begin{align}
 \label{eq:wyhel00026}
{({\rho_{\mu}} + xI)^{-1}} & = \frac{2}{1 + 2x}{(I + {\vec{v}_{\mu}}\cdot\vec{\sigma})^{-1}} \nonumber \\ 
& = \frac{2(I - {\vec{v}_{\mu}}\cdot\vec{\sigma})}{(1 + 2x)(1 - {|{\vec{v}_{\mu}}|^2})} \nonumber \\ 
& = \frac{2[(1 + 2x)I - {\vec{r}_{\mu}}\cdot\vec{\sigma}]}{{(1 + 2x)^2} - {|{\vec{r}_{\mu}}|^2}} ~,
\end{align}
with ${\vec{r}_{\mu}} = (1 + 2x){\vec{v}_{\mu}}$, and thus
\begin{equation}
 \label{eq:wyhel00027}
{\rho_{\mu}}{({\rho_{\mu}} + xI)^{-1}} = \frac{1 + 2x - {|{\vec{r}_{\mu}}|^2}}{{(1 + 2x)^2} - {|{\vec{r}_{\mu}}|^2}}I + 
 \frac{2x}{{(1 + 2x)^2} - {|{\vec{r}_{\mu}}|^2}}({\vec{r}_{\mu}}\cdot\vec{\sigma})  ~.
\end{equation}
Substituting the previous result into Eq.~\eqref{eq:wyhel00025} and performing the calculation of 
both integrals, we finally obtain
\begin{equation}
 \label{eq:wyhel00029}
\sqrt{\rho_{\mu}} = \frac{1}{2\sqrt{2}}[{c_+}I + {c_-}({\hat{r}_{\mu}}
\cdot\vec{\sigma})] ~,
\end{equation}
where
\begin{equation}
 \label{eq:wyhel0002902}
 {c^{\pm}_{\mu}} := \sqrt{1 + {|{\vec{r}_{\mu}}|}} \pm 
\sqrt{1 - {|{\vec{r}_{\mu}}|}} ~.
\end{equation}

As pointed out in the main text, the Hellinger angle is determined by the equation $\cos[\mathcal{L}({\rho_0},{\rho_{\varphi}})] 
= \text{Tr}(\sqrt{\rho_0}\sqrt{\rho_{\varphi}})$. Starting from the result indicated in Eq.~\eqref{eq:wyhel00029}, we conclude
\begin{equation}
 \label{eq:wyhel00038}
\cos[\mathcal{L}({\rho_0},{\rho_{\varphi}})] = \frac{1}{4}\left[{c_0^+}{c_{\varphi}^+} + {c_0^-}{c_{\varphi}^-}
({\hat{r}_{\varphi}}\cdot{\hat{r}_0})\right] ~.
\end{equation}
Analogously, the Wigner-Yanase skew information $\mathcal{I}({\rho_{\varphi}},{K_{\varphi}}) = -(1/2)\text{Tr}
([\sqrt{\rho_{\varphi}},{K_{\varphi}}]^2)$ also depends on the square root of the density operator. Considering the 
local observable ${K_{\varphi}} = \varpi(\alpha{I} + {\hat{n}_{\varphi}}\cdot\vec{\sigma})$ it is possible to prove that 
\begin{equation}
 \label{eq:wyhel0004601}
[\sqrt{\rho_{\varphi}},{K_{\varphi}}] = i\frac{\varpi{c_{\varphi}^-}}{\sqrt{2}}({\hat{r}_{\varphi}}\times{\hat{n}_{\varphi}})\cdot\vec{\sigma}
\end{equation}
and also
\begin{equation}
 \label{eq:wyhel0004602}
[\sqrt{\rho_{\varphi}},{K_{\varphi}}]^2 = -\frac{1}{2}{(\varpi{c_{\varphi}^-})^2}{|{\hat{r}_{\varphi}}\times{\hat{n}_{\varphi}}|^2}I ~,
\end{equation}
where we used Eq.~\eqref{eq:wyhel00029} in order to write
\begin{multline}
 \label{eq:wyhel00044}
\sqrt{\rho_{\varphi}}{K_{\varphi}} = \frac{\varpi}{2\sqrt{2}}[\alpha{c_{\varphi}^+} + {c_{\varphi}^-}
({\hat{r}_{\varphi}}\cdot{\hat{n}_{\varphi}})]I + \\ + \frac{\varpi}{2\sqrt{2}}[{c_{\varphi}^+}{\hat{n}_{\varphi}} + \alpha{c_{\varphi}^-}
{\hat{r}_{\varphi}} + i{c_{\varphi}^-}({\hat{r}_{\varphi}}\times{\hat{n}_{\varphi}})]\cdot\vec{\sigma}
\end{multline}
and
\begin{multline}
 \label{eq:wyhel00045}
{K_{\varphi}}\sqrt{\rho_{\varphi}} = \frac{\varpi}{2\sqrt{2}}[\alpha{c_{\varphi}^+} + {c_{\varphi}^-}
({\hat{r}_{\varphi}}\cdot{\hat{n}_{\varphi}})]I + \\ + \frac{\varpi}{2\sqrt{2}}[{c_{\varphi}^+}{\hat{n}_{\varphi}} + \alpha{c_{\varphi}^-}
{\hat{r}_{\varphi}} - i{c_{\varphi}^-}({\hat{r}_{\varphi}}\times{\hat{n}_{\varphi}})]\cdot\vec{\sigma} ~.
\end{multline}
Therefore, the Wigner-Yanase skew-information is given by
\begin{equation}
 \label{eq:wyhel00047}
\mathcal{I}({\rho_{\varphi}},{K_{\varphi}}) = \frac{1}{2}{(\varpi{c_{\varphi}^-})^2}{|{\hat{r}_{\varphi}}
\times{\hat{n}_{\varphi}}|^2} ~.
\end{equation}
Remember that the quantum system evolves under an unitary transformation which does not change the 
absolute value of Bloch sphere radius, i.e., $|{\vec{r}_{\varphi}}| = |{\vec{r}_0}| = {r_0}$. Therefore, since 
${c_{\varphi}^{\pm}} = {c_0^{\pm}}$ and defining ${\xi_{\pm}} = 1 \pm \sqrt{1 - {r_0^2}}$, the cosine of the 
Hellinger angle and Wigner-Yanase skew information becomes, respectively,
\begin{equation}
 \label{eq:wyhel00042}
\cos[\mathcal{L}({\rho_0},{\rho_{\varphi}})] = \frac{1}{2}[{\xi_+} + {\xi_-}({\hat{r}_{\varphi}}\cdot{\hat{r}_0})]
\end{equation}
and
\begin{equation}
 \label{eq:wyhel00043}
 \mathcal{I}({\rho_{\varphi}},{K_{\varphi}}) = {\varpi^2}{\xi_-}{|{\hat{r}_{\varphi}}\times{\hat{n}_{\varphi}}|^2} ~.
\end{equation}


\bibliographystyle{apsrev4-1}

\begin{thebibliography}{82}
\bibitem [{Our()}]{Ourspeedlimit}%
  \BibitemOpen
  \href@noop {} {\bibinfo  {journal} {To be published elsewhere}\ }\BibitemShut
  {NoStop}%
\end{thebibliography}

%

\end{document}